\documentclass[letterpaper]{JHEP3}

\newcommand{\beq}{\begin{equation}}
\newcommand{\eeq}{\end{equation}}
\newcommand{\cA}{{\cal A}}
\newcommand{\cAb}{{\overline{\cal A}}}
\newcommand{\cF}{{\cal F}}
\newcommand{\cFb}{{\overline{\cal F}}}
\newcommand{\cD}{{\cal D}}

\newcommand{\cDb}{{\overline{\cal D}}}
\newcommand{\cQ}{{\cal Q}}
\newcommand{\cU}{{\cal U}}
\newcommand{\cUb}{{\overline{\cal U}}} 
\newcommand{\KD}{{K\"{a}hler-Dirac }}
\newcommand{\cN}{{\cal N}}
\newcommand{\Tr}{{\rm Tr\;}}
\newcommand{\bx}{{\bf x}}
\newcommand{\bmu}{{\boldsymbol \mu}}
\newcommand{\bnu}{{\boldsymbol \nu}}
\newcommand{\blambda}{{\boldsymbol \lambda}}

\usepackage{graphicx}
\usepackage{amsmath}
\usepackage{amsfonts}
\usepackage{epsfig}

\title{Topological gravity on the lattice}

\author{Simon Catterall \\
Department of Physics, Syracuse University, Syracuse, NY13244, USA \\
E-mail: \email{smc@phy.syr.edu}
}

\preprint{}

\date{}

\abstract{
In this paper we show that a particular
twist of $\cN=4$ super Yang-Mills in three dimensions with
gauge group $SU(2)$
possesses a set of classical vacua corresponding
to the space of flat connections of the {\it complexified} gauge
group $SL(2,C)$. The theory also contains
a set of topological observables
corresponding to Wilson loops wrapping non-trivial cycles of
the base manifold.
This moduli space and set of topological observables
is shared with the Chern Simons formulation of three dimensional
gravity and we hence conjecture that the Yang-Mills theory
gives an equivalent description of the gravitational theory.
Unlike the Chern Simons formulation the twisted
Yang-Mills theory possesses a supersymmetric and gauge
invariant lattice
construction which then provides
a possible non-perturbative definition
of three dimensional gravity.
}

\begin{document}

%-----------------------------------------------------------------
%
\section{Introduction}

The construction of topological field theories from twisted versions
of supersymmetric theories has a long history going back to Donaldson-Witten
theory and its realization in terms of a twist
of $\cN=2$ super Yang-Mills \cite{Witten:1988ze}.
More recently these twisted theories have received a lot of attention
as the starting point for construction of lattice theories
with exact supersymmetry \cite{Catterall:2009it}. While these theories are
formulated using an arbitrary background metric they typically contain a set
of topological operators whose expectation values are independent of this
metric and correspond to integrals over the
moduli space of the theory. 

In this paper we show that a particular
twist of $\cN=4$ super Yang-Mills in three dimensions 
with gauge group $SU(2)$ possesses a moduli space corresponding
to the space of flat connections of the {\it complexified} gauge
group $SL(2,C)$. This moduli space
is shared with the Chern-Simons formulation of three dimensional
Euclidean gravity \cite{Witten:1988hc,Witten:1989sx}. 
Furthermore, as for the Chern Simons theory,
the twisted theory contains a natural set of
topological observables, corresponding to Wilson loops of the
complexified gauge field,
whose expectation values are metric independent.

Unlike the Chern Simons action, the twisted
theory has an action that is bounded from below and
possesses only a compact $SU(2)$ gauge symmetry. The path
integral defining the quantum theory
is thus well defined.
Indeed a lattice theory may be constructed
which targets this twisted theory in the naive
continuum limit while preserving both gauge invariance
and the scalar BRST supercharge that
arises after twisting. This lattice theory was first
constructed using orbifold methods in \cite{Cohen:2003qw}
and can be used to provide 
a non-perturbative definition of the
supersymmetric Yang-Mills theory in flat space. 
Furthermore, as for the continuum theory, the presence
of an exact BRST symmetry means that expectation values
of topological observables can be shown to be independent 
of the coupling constant and the lattice spacing.

In this paper we point out
that there is strong evidence to
indicate that this
lattice theory may also serve as a non-perturbative regulator for
the Chern Simons formulation of three dimensional gravity. 

The outline of this paper is as follows;
first we describe the continuum twisted theory with gauge group
$SU(N)$ formulated on an arbitrary background geometry. We then
construct a lattice model which targets the flat
space theory in the naive
continuum limit and which is both gauge invariant and
invariant under a twisted scalar supercharge. We then review
the usual
construction of three dimensional gravity as a Chern-Simons theory.  We
show that the moduli space
of this Chern-Simons theory is precisely
the same as that of the twisted Yang-Mills model with gauge group
$SU(2)$. Furthermore, both theories possess a set of topological
operators corresponding to Wilson loops of the complexified
theory which wrap the boundaries, whose
vacuum expectation
values are independent of any background
metric. Indeed, we argue that
they may be computed exactly in the
associated lattice theory.

%-------------------------------------------------------------ta----
\section{Twisted $\cN=4$ gauge theory in three dimensions}
\label{gauge}

The twist of $\cN=4$ super Yang-Mills that we are interested in can be 
most succinctly written in the form 
\beq
S=\beta(S_1+S_2)
\eeq
where 
\begin{eqnarray}
S_1&=&\cQ\int d^3x\;\sqrt{h} \left(\chi^{\mu\nu}\cF_{\mu\nu}+
\eta\left[\cDb^\mu,\cD_\mu\right]+\frac{1}{2}\eta d\right)\nonumber\\
S_2&=&\int d^3x\; \theta_{\mu\nu\lambda}\cDb^\lambda \chi^{\mu\nu}
\label{action1}
\end{eqnarray}
The fermions comprise one \KD multiplet of p-form fields 
$(\eta,\psi_\mu,\chi_{\mu\nu},\theta_{\mu\nu\lambda})$\footnote{It is common in the
continuum literature to replace the 2 and 3 form fields in these
expressions by
their Hodge duals; a second vector $\hat{\psi}_\mu$ and scalar $\hat{\eta}$ see,
for example \cite{Blau:1996bx}} where in three
dimensions $p=0\ldots 3$. Notice that a single \KD field possesses eight single
component fields as expected for a theory with $\cN=4$ supersymmetry in
three dimensions.
The imaginary parts of the
complex gauge field $\cA_\mu\;\mu=1\ldots 3$ appearing in this construction
yield the three scalar fields of the conventional (untwisted) theory. 
The scalar nilpotent
supersymmetry $\cQ$ acts on the twisted fields as follows
\begin{eqnarray}
\cQ \cA_\mu&=&\psi_\mu\nonumber\\
\cQ \cAb^\mu&=&0\nonumber\\
\cQ \psi_\mu&=&0\nonumber\\
\cQ \chi^{\mu\nu}&=&\cFb^{\mu\nu}\nonumber\\
\cQ \eta&=&d\nonumber\\
\cQ d&=&0\nonumber\\
\cQ \theta_{\mu\nu\lambda}&=&0
\label{susy}
\end{eqnarray}
where the background metric $h_{\mu\nu}$ is used to raise and
lower indices in the usual manner
$\chi^{\mu\nu}=h^{\mu\alpha}h^{\nu\beta}\chi_{\alpha\beta}$
and is a $\cQ$-singlet. The topological character of the
theory follows from the $\cQ$-exact sructure of $S_1$ together
with the explicit metric indepdence of $S_2$ which can be
recognized as the integral of a 3-form since
$\theta_{\mu\nu\lambda}=\theta\epsilon_{\mu\nu\lambda}$. 
Furthermore, $S_1$ is clearly
$\cQ$-invariant by virtue of the nilpotent property of
$\cQ$ while the supersymmetric invariance of the $\cQ$-closed term
relies on the
Bianchi identity
\beq
\epsilon^{\lambda\mu\nu}\cD_\lambda \cF_{\mu\nu}=0
\label{bianchi}
\eeq
The complex covariant derivatives appearing in these expressions are defined
by
\begin{eqnarray}
\cD_\mu&=&\partial_\mu+\cA_\mu=\partial_\mu+A_\mu+iB_\mu\nonumber\\
\cDb^\mu&=&\partial^\mu+\cAb^\mu=\partial^\mu+A^\mu-iB^\mu 
\end{eqnarray}
while all fields take values in the adjoint
representation of $SU(N)$\footnote{The generators are taken to be {\it
anti-hermitian} matrices satisfying $\Tr (T^aT^b)=-\delta^{ab}$}.
It should be noted that despite the appearance of a complexified
connection and field strength the theory possesses only the usual
$SU(N)$ gauge invariance corresponding to the real part of the gauge field.
The structure of this
twisted theory is similar to that of the Marcus twist of
$\cN=4$ super Yang-Mills in four dimensions \cite{Marcus,Unsal:2006qp,
Catterall:2007kn} 
which plays an important role in the Geometric-Langlands 
program \cite{Kapustin:2006pk}. 

Doing the $\cQ$-variation and integrating out the field $d$ yields
\beq
S=\frac{1}{g^2}\int d^3 x\;\sqrt{h}\left(L_1+L_2\right)\eeq
where
\begin{eqnarray}
L_1&=&\Tr \left(
-\cFb^{\mu\nu}\cF_{\mu\nu}+\frac{1}{2}[ \cDb^\mu, \cD_\mu]^2)\right)\nonumber\\
L_2&=&\Tr \left(
-\chi^{\mu\nu}\cD_{\left[\mu\right.}\psi_{\left.\nu\right]}-
\psi_\mu\cDb^\mu\eta -\theta_{\mu\nu\lambda}\cDb^{\left[\lambda\right.}
\chi^{\left.\mu\nu\right]}\right)
\label{action}
\end{eqnarray}
The terms appearing in $L_1$ can then be written
\begin{eqnarray}
\cFb^{\mu\nu}\cF_{\mu\nu}&=&(F_{\mu\nu}-[B_\mu,B_\nu])
                            (F^{\mu\nu}-[B^\mu,B^\nu])+
(D_{\left[\mu\right.}B_{\left.\nu\right]})
(D^{\left[\mu\right.}B^{\left.\nu\right]})\nonumber\\
\frac{1}{2}\left[\cDb^\mu,\cD_\mu\right]^2 &=& -2\left(D^\mu B_\mu\right)^2
\end{eqnarray}
where $F_{\mu\nu}$ and $D_\mu$ denote the usual field strength and
covariant derivative depending on the real part of the connection $A_\mu$.
The classical vacua of this theory correspond to solutions of
the equations
\begin{eqnarray}
F_{\mu\nu}-[B_\mu,B_\nu]&=&0\nonumber\\
D_{\left[\mu\right.}B_{\left.\nu\right]}&=&0\nonumber\\
D^\mu B_\mu&=&0
\label{moduli}
\end{eqnarray}
The same moduli space arises in
the study of the Marcus twist of four dimensional
$\cN=4$ Yang-Mills where it is argued to correspond to
the space of flat {\it complexified} connections modulo {\it complex} gauge
transformations. A simple way to understand this is to recognize
that the additional term $D^\mu B_\mu=0$ appearing in the vacuum
equations~\ref{moduli} resembles a partial gauge fixing   
of a theory with a complexified gauge invariance 
and associated gauge fields $A_\mu$ and $B_\mu$
down to a theory
possessing just the usual $SU(N)$ 
gauge invariance and gauge field $A_\mu$.

More specifically,
Marcus showed in \cite{Marcus} that the solutions of eqns.~\ref{moduli}
modulo $SU(N)$ gauge transformations are in one to one
correspondence with the space of
flat complexified $SU(N)$ connections modulo complex gauge
transformations. These arguments should hold in
the three dimensional case too.

The topological character of the theory
then guarantees that any $\cQ$-invariant observable such as the
partition function can be evaluated {\it exactly} by considering only
Gaussian fluctuations about such vacuum configurations. Furthermore, 
it is easy to see from eqn.~\ref{action1}
that the energy momentum tensor of this
theory is $\cQ$-exact rendering the expectation values of such
topological observables independent of smooth deformations
of the background metric $h_{\mu\nu}(x)$.

Returning to the bosonic action and integrating 
by parts we find that the term linear in $F_{\mu\nu}$ cancels
and the contribution of $L_1$ reads
\beq
L_1=\Tr \left(F_{\mu\nu}F^{\mu\nu}+
2B^\mu D^\nu D_\nu B_\mu-[B_\mu,B_\nu][B^\mu,B^\nu]-2R^{\mu\nu}B_\mu
B_\nu\right)
\label{final}\eeq
where $R_{\mu\nu}$ is the background Ricci tensor.
Thus the bosonic theory possesses the usual
Yang-Mills field strength for a real
$SU(N)$ gauge field together with
three vectors arising from the imaginary part of the connection. 
In flat space the theory is fully
equivalent to the usual $\cN=4$ theory as this twisting operation can
be regarded merely as an exotic change of variables -- the 
vectors can regarded as scalar fields since their kinetic term is
simply the usual scalar Laplacian and the \KD action is
equivalent to the Dirac action for four degenerate
Majorana spinors. 

Notice that the fact that the complexified
connection $\cAb_\mu$ is a $\cQ$-singlet
allows us to trivially construct a class of topological observables
corresponding to the trace of an associated Wilson loop 
around a non-trivial cycle $\gamma$ in the background
space
\beq 
O(\gamma)={\cal P} e^{ \int_\gamma \cAb^\mu .dx_\mu}\eeq 

As we will argue later the specialization of this model to the case
of $SU(2)$ is particularly interesting as 
the resulting moduli space coincides with that arising in
three dimensional Chern Simons gravity.

\section{Lattice construction}

The twisted theory described in the previous section may be discretized
using the techniques developed in \cite{Catterall:2007kn,Damgaard:2008pa,Damgaard:2007be}. 
The resultant lattice theories
have the merit of preserving both gauge invariance and the scalar
component of the twisted
supersymmetry. 
Here we show how to derive this lattice theory by direct discretization
of the continuum twisted theory. We will start by assuming
that the continuum theory
is formulated in flat (Euclidean) space 
with metric $h_{\mu\nu}=\delta_{\mu\nu}$. 
This guarantees that the twisted theory is completely equivalent
to the usual Yang-Mills theory and hence that the resulting lattice
models target the usual continuum theory in the continuum limit.

Furthermore, in the case of topological observables the choice of
metric is unimportant and hence the lattice theory we construct will
yield expectation values for topological operators which depend
only on the topology of the lattice and not on the coupling,
lattice spacing or the fact that we started by discretization of
a theory in a flat background. 

The transition to the lattice from the continuum theory 
requires a number of steps. The first, and most important,
is to replace the continuum complex gauge field $\cA_\mu(x)$ at every
point
by an appropriate complexified Wilson
link $\cU_\mu(\bx)=e^{\cA_\mu(\bx)},\mu=1\ldots 3$. 
These lattice fields 
are taken to be
associated with unit length vectors in the coordinate
directions $\bmu$ in an abstract
three dimensional hypercubic lattice. 
By supersymmetry the fermion
fields $\psi_\mu(\bx),\mu=1\ldots 3$ lie on the same oriented
link as their
bosonic superpartners running from $\bx\to\bx+\bmu$. In contrast the scalar
fermion $\eta(\bx)$ is associated with the site $\bx$
of the lattice and the
tensor fermions $\chi^{\mu\nu}(\bx),\mu<\nu =1\ldots 3$ 
with a set of diagonal face links
running from $\bx+\bmu+\bnu\to \bx$. The final 3 form field
$\theta_{\mu\nu\lambda}(\bx)$ is then naturally placed on the
body diagonal running from $\bx\to \bx+\bmu+\bnu+\blambda$. 
The construction then posits that
all link fields transform as bifundamental
fields under gauge transformations
\begin{eqnarray}
\eta(\bx)&\to& G(\bx)\eta(\bx) G^\dagger(\bx)\nonumber\\
\psi_\mu(\bx)&\to& G(\bx)\psi_\mu(\bx) G(\bx+\bmu)\nonumber\\
\chi^{\mu\nu}(\bx)&\to&G(\bx+\bmu+\bnu)\chi^{\mu\nu}(\bx)G^\dagger(\bx)\nonumber\\
\cU_\mu(\bx)&\to&G(\bx)\cU_\mu(\bx)G^\dagger(\bx+\bmu)\nonumber\\
\cUb^\mu(\bx)&\to&G(\bx+\bmu)\cUb^\mu(\bx)G^\dagger(\bx)
\label{gaugetrans}
\end{eqnarray}
Notice that
we can keep track of the orientation of the lattice field by following
its continuum index structure -- upper index fields are placed on
negatively orientated links, lower index fields live on positively
oriented links. 

The action of the scalar supersymmetry on these fields is given by the
continuum expression in eqn.~\ref{susy} with the one modification
that the continuum field $\cA_\mu(x)$ is replaced with the
Wilson link $\cU_\mu(x)$ and the lattice field strength being defined
as $\cF_{\mu\nu}=D^{(+)}_\mu U_\nu$.
The supersymmetric and gauge invariant
lattice action which corresponds to
eqn.~\ref{action} then takes a very similar form
to its continuum counterpart
\begin{eqnarray}
S_1&=&\cQ\sum_{\bx} \left(\chi^{\mu\nu}\cF_{\mu\nu}+
\eta \left[ \cDb^{(-)\mu}\cU_\mu \right]+\frac{1}{2}\eta d\right)\nonumber\\
S_2&=&\sum_{\bx} \theta_{\mu\nu\lambda}\cDb^{(+)\lambda} \chi^{\mu\nu}
\label{act}
\end{eqnarray}
The covariant difference operators appearing in these expressions are
defined by 
\begin{eqnarray}
\cD^{(+)}_\mu f_\nu(\bx)&=&
\cU_\mu(\bx)f_\nu(\bx+\bmu)-f_\nu(\bx)\cU_\mu(\bx+\bnu)\nonumber\\
\cDb^{(-)\mu} f_\mu(\bx)&=&
f_\mu(\bx)\cUb^\mu(\bx)-\cUb^\mu(\bx-\bmu)f_\mu(\bx-\bmu)
\end{eqnarray}
These expressions are determined by the twin requirements that they reduce
to the corresponding continuum results for the adjoint covariant derivative
in the naive continuum
limit $\cU_\mu\to 1+\cA_\mu+\ldots$ and that they transform
under gauge transformations like the corresponding lattice
link field carrying the same indices. This allows the terms in the
action to correspond to gauge invariant closed loops on the lattice.
Similarly the difference operator appearing in $S_2$ takes the form
\beq
\cDb^{(+)\lambda}\chi^{\mu\nu}(\bx)=
\chi^{\mu\nu}(\bx+\blambda)\cUb^\lambda(\bx)-
\cUb^\lambda(\bx+\bmu+\bnu)\chi^{\mu\nu}(\bx)
\eeq
This definition allows the lattice term corresponding
to $S_2$ to be both gauge invariant and supersymmetric -- the latter
property holding because of the remarkable property that the
lattice field strength satisfies an exact Bianchi identity as for
the continuum \cite{Catterall:2007kn}. 
The action can also be
shown to be free of fermion doubling problems -- see the
discussion in \cite{Catterall:2007kn}.

As in the continuum,
the presence of an exact $\cQ$-symmetry allows the definition of a 
class of supersymmetric Wilson loop corresponding to the trace of
the product of $\cUb_\mu$ links around a closed loop in the lattice.
\beq
O=\prod_{t=1}^T \cUb^t(\bx)\eeq
The vacuum expectation value of these operators can be computed
exactly by restriction to the moduli space of theory and can
probe only topological features of the background space.

\section{Chern-Simons formulation of three dimensional gravity}
\label{spinstuff}

The twisted model we have discussed appears on the face of
it to have little connection
to gravity. However it has been known for a long time that
three dimensional gravity can be reformulated in the language of
gauge theory \cite{Witten:1988hc,Witten:1989sx}. For
a review of three dimensional gravity see \cite{Carlip:1995zj}. 
The construction employs
a Chern-Simons action and in Euclidean space the local symmetry
corresponds to the group $SO(3,1)\sim SL(2,C)$ - the complexified
$SU(2)$ group. Furthermore the resulting theory is topological
and determined by the moduli space of flat $SL(2,C)$ connections
hinting at a close connection to the twisted Yang-Mills theory described
above. We now summarize this theory which will allow us to re-interpret
the fields in the two color Yang-Mills theory described
earlier as geometrical objects in
a gravitational theory.

Consider the following three dimensional Chern-Simons action
\beq
\int d^3 x\epsilon_{\mu\nu\lambda}{\rm \hat{Tr}}\left(A_\mu F_{\nu\lambda}
-\frac{1}{3}A_\mu \left[A_\nu , A_\lambda\right]\right)
\label{CS}\eeq
Furthermore, assume that the gauge field $A_\mu$ takes values
in the adjoint representation of the group $SO(3,1)$. A convenient
representation for the six generators of the Lie algebra of
this group is then given by commutators
of the (Lorentzian) Dirac matrices 
$\gamma^{AB}=\frac{1}{4}\left[\gamma^A,\gamma^B\right]$ where
$\left(\gamma^a\right)^\dagger=\gamma^a\;a=1\ldots 3$ and
$\left(\gamma^4\right)^\dagger=-\gamma^4$
This yields an expression for the gauge field of the form
\beq
A_\mu=\sum_{A<B}A_\mu^{AB}\gamma^{AB}\quad A,B=1\ldots 4\eeq
Finally, the group indices are contracted using the invariant tensor
$\epsilon_{ABCD}$ corresponding to a trace of the form
\beq
{\rm \hat{Tr}}(X)={\rm Tr}(\gamma_5 X)\eeq
Notice that this way of contracting the group indices
differs from the simple trace
that appears in the twisted Yang-Mills theory
considered in the previous section. 
To see explicitly that the resulting theory is just three dimensional
gravity we decompose the gauge field and field strength in terms of
an $SO(3)$ subgroup
\begin{eqnarray}
A_\mu&=&\sum_{a<b}\omega_\mu^{ab}\gamma^{ab}+\frac{1}{l}e_\mu^{a}\gamma^{4a}\quad
a,b=1\ldots 3\nonumber\\ 
F_{\mu\nu}^{ab}&=&\sum_{a<b} \left(R_{\mu\nu}^{ab}+\frac{1}{l^2}e_{\left[\mu\right.}^a
e_{\left.\nu\right]}^b\right)\gamma^{ab}\nonumber\\
F_{\mu\nu}^a&=&\sum_a D_{\left[\mu\right.}e_{\left.\nu\right]}^a
\end{eqnarray}
The covariant derivative appearing in the field strength contains
just the $SO(3)$ gauge field $\omega_\mu$ and we have
introduced a explicit length scale $l$ into the definition of
the gauge fields $e_\mu$.
After substituting into the Chern-Simons
action one recognizes that it corresponds to 
three dimensional Einstein-Hilbert gravity including
a cosmological constant and written in the
first order tetrad-Palatini formalism \cite{Witten:1988hc}.
\beq
S_{\rm EH}=\frac{1}{l}\int \epsilon^{\mu\nu\lambda} \epsilon_{abc}\left(
e^a_\mu R^{bc}_{\mu\nu}-\frac{1}{3l^2}e^a_\mu e^b_\nu e^c_\lambda\right)
\label{palatini}\eeq
with $\omega_\mu$ and $e_\mu$ corresponding to the spin connection
and dreibein and $1/l^2$ playing the role of a cosmological
constant.
To see that this theory is classically equivalent to the usual metric
theory of gravity one merely has to notice that the
classical equations of the Chern-Simons theory require that 
the $SO(3,1)$ field strength vanish $F_{\mu\nu}^{AB}=0$
which sets both the torsion $T=D_{\left[\mu \right.} e_{\left. \nu\right]}$ 
to zero and requires a constant $SO(3)$ curvature $R_{\mu\nu}$ equal
to $-1/l^2$. Such a solution corresponds (at least locally)
to hyperbolic three space $H^3\sim SO(3,1)/SO(3)$.

Finally one can show that the theory restricted to
this space of flat connections is also invariant under
diffeomorphisms \cite{Witten:1988hc}. 
This result follows from the fact that one can express
a general coordinate transformation on $A_\mu$
with parameter $-\xi^\nu$ as
a gauge transformation with parameter $\xi^\mu A_\mu$ plus a term
which vanishes on flat connections.
\beq
\delta A_\mu^\xi=-D_\mu (\xi^\nu A_\nu)-\xi^\nu F_{\mu\nu}\eeq
Furthermore,
the fact that this action is
explicitly independent of any background metric ensures that the theory
is topological. The classical solutions correspond to the space
of flat $SO(3,1)$ connections up to $SO(3,1)$ gauge transformations.
Since $SO(3,1)\sim SL(2,C)$ this is equivalent to the moduli space
appearing earlier in the twisted model. Indeed, the topological observables
in that theory, corresponding to Wilson loops of the complexified
gauge field $\cAb_\mu$,
map into the natural topological observables of the Chern-Simons
theory - Wilson loops of an $SO(3,1)$ connection\footnote{This
mapping between connections is manifest if the Weyl basis is used
for the Dirac matrices which leads to generators proportional to
$\sigma$ and $i\sigma$}. 

Indeed, an explicit connection between
the Yang-Mills theory and the gravity
theory would identify the imaginary parts
of the $SL(2,C)$ connection - the field $B_\mu$ occurring
in the Yang-Mills theory - with the
matrix valued field
$e_\mu$ occurring the tetrad-Palatini action. 
Notice that the field $B_\mu$ transforms in the adjoint representation
of the $SU(2)$ gauge group which translates in the
gravitational theory to the statement that
the dreibein $e_\mu^a$ transforms 
as a vector under local Lorentz
transformations as it should.

These considerations together with the
equivalence of the topological sectors
of these two theories leads us
to conjecture that the twisted two color Yang-Mills gives an
alternative representation of the gravity theory. 
Furthermore, this alternative representation has some
advantages -- the path integral is now well defined and indeed
may be given a non-perturbative definition as the appropriate
limit of a gauge and supersymmetric invariant lattice model.

\section{Discussion}

In this paper we have shown how a twist of $\cN=4$ super Yang-Mills
theory in three dimensions
with gauge group $SU(2)$ shares both a moduli space and
a set of topological observables in common with
the Chern Simons formulation of three dimensional Euclidean
gravity.
Indeed, on this basis we conjecture that the topological sector
of the twisted Yang-Mills theory is equivalent to the
Chern Simons theory. This is particularly interesting in the light
of the fact that this twisted theory may be discretized while
maintaining the BRST symmetry of the twisted model and hence
its topological properties. Indeed, the lattice
model can be thought of as supplying a
rigorous non-perturbative definition of the twisted
Yang-Mills model. By these arguments it thus also
defines a lattice theory
of topological gravity. 
This lattice theory
may  be simulated using standard Monte Carlo techniques similar
to those reported from initial simulations of its four dimensional
cousin \cite{Catterall:2008dv}. 

It has been known for quite some time that topological gravity
in odd dimensions can be formulated as a Chern Simons theory.
In particular such a formulation exists in five dimensions
\cite{Chamseddine:1989nu,Zanelli:2005sa}. It would
be very interesting to see whether it is
possible to construct a twisted Yang-Mills theory in five dimensions
which targets the same moduli space as
this Chern Simons theory.

\acknowledgments The author is supported in part by DOE grant
DE-FG02-85ER40237. The author would like to acknowledge Poul
Damgaard and Daniel Ferrante for useful discussions.

%-----------------------------------------------------------------
%
\bibliographystyle{JHEP}
\bibliography{AdS3D}
%
%-----------------------------------------------------------------

\end{document}